\documentstyle[aps,multicol,epsf]{revtex}

\begin{document}

\draft

\title{Liouville Field Theory of Fluctuating Loops}
 
\author{Jan\'{e}  Kondev}

\address{Department of Physics, Brown University, Providence, Rhode 
Island 02912-1843}

\date{March 12, 1997}

\maketitle

\widetext

\begin{abstract}
Effective  field theories of two-dimensional lattice models of 
fluctuating loops are constructed by mapping them onto
random surfaces whose large scale fluctuations are described by 
a Liouville field theory. This provides a {\em geometrical}
view of conformal invariance in two-dimensional critical phenomena 
and a  method for calculating critical properties of loop  
models exactly.
As an application of the method, the conformal charge and 
critical exponents for two mutually excluding Hamiltonian 
walks on the square lattice are calculated.  
\end{abstract}

\vspace{-1mm}


\vspace{-5mm}

\begin{multicols}{2}

\narrowtext

The study of critical phenomena in two dimensions has 
led to a  rather remarkable interplay between  statistical  
mechanics and field  theory.  The picture that has emerged 
portrays  critical  fluctuations in a variety  of systems  as  
being described by a  conformal field theory (CFT), 
i.e., a quantum field theory invariant 
under {\em local} scale (conformal) transformations. This view 
has proven very fruitful as it has answered many   
questions pertaining to two-dimensional phase transitions. 
For example, by studying the 
representations of the Virasoro algebra, which is  the algebra of 
conformal transformations, we have come to  understand 
why critical exponents  in two-dimensions  are typically 
{\em rational}  numbers \cite{bpz}. Lists of possible 
critical exponents appear here in very much the same way one 
discovers the quantization of angular momentum by studying 
the representations of the algebra  of rotations. 

Progress in understanding two-dimensional critical phenomena using 
CFT's has largely come about by classifying  
these field theories and associating them with the scaling 
limits of specific  microscopic  lattice  models
\cite{bpz,cardyrev}. 
Still a fundamental  practical question remains: Given a 
lattice model how does one go about {\em constructing} a
conformal field  theory of its  scaling limit? 
It is this question, limited in scope to lattice models of fluctuating 
loops \cite{warnaar}, which is addressed here. Since most  
canonical two-dimensional models can be recast as  loop models 
\cite{baxterbook},   the results  we find are   
general in nature, and   provide a  {\em geometrical} view of 
criticality and conformal invariance in  two-dimensions. 
In particular, we 
find that CFT's of loop models can be constructed explicitly 
via a  mapping of the loop model to a model of a fluctuating surface. 
The basic idea is to think of loops  as   
{\em contour lines} of a random  surface;    
the field theory which  describes the large-scale 
fluctuations of this surface is a conformally invariant 
{\em Liouville field  theory}.

Liouville field theory provides a general framework for 
calculating correlation functions in CFT's; this is the 
well known {\em Coulomb gas} representation pioneered  by 
Dotsenko and Fateev \cite{dots}. 
The central result of this paper is the  
discovery that different terms in the Liouville 
action, some of which were originally  introduced on formal grounds, 
have a concrete {\em geometrical} interpretation in 
loop models. Furthermore, the coupling constants, 
most notably  the effective stiffness of the surface $K$,  
are fixed by geometry, thus   providing a method for    
calculating  critical properties of loop models {\em exactly}.
To illustrate this we 
calculate the 
conformal charge and critical exponents for a new loop model 
defined on the square lattice. For  vanishing loop weights  
this model defines a new universality 
class of  Hamiltonian walks; these are self-avoiding 
random walks that visit all the sites of the lattice. Hamiltonian 
walks have been used in the past to  model  configurational 
statistics of polymer melts \cite{clozeiux}. 

\paragraph{Loop models}

Given a two-dimensional lattice ${\cal L}$ 
a loop  configuration ${\cal G}$ is defined by a set of loops, which
are closed self-avoiding 
walks  along the bonds of ${\cal L}$.
The partition function is  
$Z=\sum_{\cal G}\exp[-\beta H({\cal G})]$, where $\beta H({\cal G})$
is the reduced loop-Hamiltonian. 
 
As remarked in the introduction, 
many two-dimensional models can be recast as loop models. 
For example, in  the Potts model one arrives at a loop model 
description by considering the graphs generated by the 
high-temperature expansion of the partition 
function \cite{baxterbook}; in the  $O(n)$ model, 
the partition function takes the form of a loop model once 
integration over the spin degrees of freedom is
carried out \cite{nienrev}. 
Here we focus our attention on  {\em fully packed} loop 
(FPL) models whose allowed configurations ${\cal G}$ satisfy 
the constraint that {\em every} vertex of the lattice is 
visited by a loop, and 
\begin{equation} 
\label{hamilton}
    \beta H({\cal G}) =  \sum_{i=1}^{n_f} \mu_i N_i({\cal G}) \; .
\end{equation}
is the loop Hamiltonian. 
Here  we have allowed for $n_f$ flavors of loops; 
$n_i=\exp(-\mu_i)$  is the {\em loop weight},  
and $N_i$ is the number of  loops of flavor $i$. 
In the allowed configurations ${\cal G}$ only 
different flavored loops can cross. An important example of an 
FPL model is provided by the critical Q-state Potts model, 
which can be mapped to an FPL model on the  
(oriented)  square lattice with  $n_f=1$ and 
$n=\sqrt{Q}$ \cite{cardyrev}. 
Recent interest in FPL models 
\cite{nienfpl,batchfpl,jkfpl,squareFPL}  has been sparked by their
close relation to the problem of Hamiltonian walks: from the 
Bethe ansatz solution of the FPL model on the honeycomb lattice
exact critical properties of Hamiltonian walks on a non-oriented 
lattice were obtained for the first time \cite{batchfpl}.  

\begin{figure}
\epsfxsize=8.5cm \epsfbox{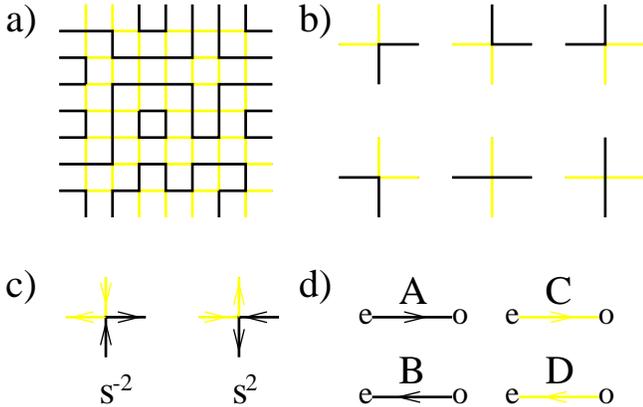}
\caption
{
\label{vconf}
The two-flavor fully packed loop (FPL${}^2$) model on the 
square lattice: a) loop 
configuration ${\cal G}$; b) allowed vertex configurations; 
c) two vertex
configurations of  oriented loops (of 24 total) and their 
weights -- $s=\exp(i\pi e_0/4)$;   
d) mapping to the coloring model -- ``e'' and ``o'' denote  
sites on the even and odd sublattice of the square lattice.   
}
\end{figure}

In order to illustrate the {\em general} relation between 
loop models and Liouville field  theories,   
we  calculate the exact conformal charge and critical 
exponents for the  {\em two-flavor} ($n_f=2$) fully packed loop 
model (FPL${}^2$) on the square lattice,  with loop weights  
$n_1=n_2=n>0$.
The two loop 
flavors of the model will be denoted black and grey, in accord 
with Fig.~\ref{vconf}a; 
the possible {\em vertex configurations} of the FPL${}^2$ model 
are shown in Fig.~\ref{vconf}b. The allowed 
loop configurations are 
the same as in the dimer-loop model \cite{clhdimer}, and the 
FPL model on the 
square lattice  \cite{squareFPL}, the difference being in the 
loop weights. For the dimer loop model $n_1=2$, $n_2=1$, while the 
FPL model  generalizes this to $n_1=n>0$, $n_2=1$.   
Obviously the FPL and FPL${}^2$ models  
share a common point: $n=1$. We will 
show later that for $n=1$ there is excellent agreement between 
the numerical 
results of  Ref.~\cite{squareFPL} and our exact expressions. 


It is instructive to look at the $n\to 0$ ($\mu\to \infty$) 
and $n\to \infty$ ($\mu\to -\infty$)
limits of the  FPL${}^2$ model. From Eq.~(\ref{hamilton}) 
we infer that in the first limit  configurations
with one black and one gray loop are selected. These 
self-avoiding walks are Hamiltonian
and they  can be thought of as a  
model of two mutually repelling species of polymers in the melt. 
Later we will show that this model defines a new universality class of 
Hamiltonian walks. In the limit  $n\to \infty$
the selected  loop configurations
maximize the number of loops, each being of minimum length four. 
In this  case the model has a finite correlation length ($\xi$), 
measured  by the average loop size,  while in the case of Hamiltonian
walks the correlation length  obviously diverges with system size. 
Later we will argue that at $n=2$ a transition 
occurs between a critical phase ($\xi\to\infty$) 
and a long range ordered one (finite $\xi$).

\paragraph{Liouville field theory}

To calculate the critical properties of a loop model we construct a 
random surface for which the loops are {\em contour lines}. 
This construction can be broken up into three steps, which are 
illustrated here for the FPL${}^2$ model. 

First we orient all the loops; 
for every loop configuration 
${\cal G}$ this gives  $2^{N_1({\cal G})+N_2({\cal G})}$ oriented-loop
configurations.  
In order to recover the proper loop weights ($n$) 
every left/right turn of the oriented black and grey loops
is assigned a weight  $\lambda=\exp(\pm {\rm i}\pi e_0/4)$; 
if the loop does not make a turn, $\lambda=1$ is the appropriate weight
(see Fig.~\ref{vconf}c).
On the square lattice the difference between the number of 
left and right  
turns that a closed loop makes is $\pm 4$, so the $\lambda$'s along a 
loop will  combine to give a weight  of 
$\exp(\pm {\rm i}\pi e_0)$, depending on the  loop orientation.  
After summing over the two possible 
orientations, the correct loop weight 
\begin{equation}
\label{e0vsn}
    n = e^{{\rm i}\pi e_0} + e^{-{\rm i}\pi e_0} = 2 \cos(\pi e_0) 
\end{equation}
is recovered. 

There is an important caveat to the above construction.
Namely, if we define the loop model on 
a cylinder, then the loops that wind around the cylinder are  weighted
incorrectly, with $n=2$, since the number
of left and right turns for these loops is  equal.
To correct  this a boundary term is introduced  
in the effective action for the 
FPL${}^2$ model --  $S_b$ in Eq.~(\ref{action}) \cite{cardyrev}.  

Second, we map configurations of oriented loops to configurations
of a microscopic height ${\bf z}=(z_1,z_2,z_3)$ 
which defines a two dimensional 
interface in five dimensions; the black and grey loops are contour 
lines of projections of ${\bf z}$ along different  directions. 
The microscopic heights 
are defined at the centers of the plaquettes of ${\cal L}$ 
in the following way. 
The difference in ${\bf z}$ between two neighboring 
plaquettes  depends on the flavor and direction of the 
bond that they share, and 
is given by one of four vectors (also referred to as colors): 
${\bf A}=(-1,1,1)$, ${\bf B}=(1,1,-1)$, 
${\bf C}=(-1,-1,-1)$, or  ${\bf D}=(1,-1,1)$; Fig.~\ref{vconf}d. 
This height mapping
is the same as the one for the four-coloring model on the 
square lattice \cite{jkprb}. 

Third, we coarse grain the microscopic height. We view the 
configurations of ${\bf z}$ as consisting of flat domains over which 
${\bf z}$ is averaged so as to obtain a coarse grained
height ${\bf h}$ \cite{jkprb}. 
The partition function of the 
loop model which incorporates only the large scale fluctuations
of ${\bf h}$ can be written as a functional integral
$\int \! {\cal D} {\bf h}  \exp(-S)$,  where $S$ is the   
effective action. $S$  can be broken up into three
terms, 
\begin{eqnarray}
\label{action}
 S    & = & S_e + S_b + S_w \nonumber \\ 
 S_e  & = & \frac{K}{2} \int \!\! d^2 {\bf x} \;
         [(\partial_1 {\bf h})^2 +(\partial_2 {\bf h})^2] \nonumber \\
 S_b  & = & \frac{{\rm i}}{4\pi} \int \!\! d^2 {\bf x} \; 
           ({\bf E}_0 \cdot {\bf h}) R                 \nonumber \\
 S_w  & = & \int \!\! d^2 {\bf x} \; w({\bf h})  \; , 
\end{eqnarray}
each having a  simple geometrical interpretation: \newline 
{\em i)} 
$S_e$ describes  the elastic fluctuations of the interface.
Its form, namely the fact that the elasticity  
is given by a single stiffness  constant $K$,  is 
fixed by the symmetries of the loop model: four-fold rotations 
of the lattice, 
translations in height space, and cyclic permutations  of the colors 
\cite{clhdimer,jkunp}.  \newline
{\em ii)} 
$S_b$ is the above mentioned boundary term that corrects the 
weights of loops that wind around a cylinder \cite{cardyrev}.
$R$ is the scalar curvature, which for 
a particular  choice of coordinates  on the cylinder 
is the difference of  two delta functions centered at the two  ends. 
The 
term $S_b$  therefore has the effect of placing {\em vertex operators} 
$\exp(\pm{\rm i}{\bf E}_0 \cdot {\bf h})$ 
at the boundaries of the cylinder.
In the Coulomb gas
representation vertex operators are associated with {\em electric}
charges, and  ${\bf E_0}$ 
is referred to as the {\em background charge} \cite{dots}. 
In the  chosen normalization for the
color-vectors  ${\bf E}_0= (-\pi e_0, 0, 0)$. \newline
{\em iii)}
$S_w$ ensures the correct weight of loops in the bulk. 
Namely, as mentioned earlier, the loops of the 
FPL${}^2$ model are {\em contours} of ${\bf h}$, and if the 
only bulk term in $S$ were $S_e$ then the two loop orientations would 
be equivalent -- $S_e$ equally weights a step up and a step down in 
height. 
This is inconsistent with $n\neq 2$ and an additional bulk term  
is necessary.  

It is important to note  that a
similar construction, leading to $S_e$ and $S_b$, has been used 
previously by many authors, and it  goes   under the name:
Coulomb gas approach to critical phenomena 
\cite{nienrev,difran}. The important difference here is 
the inclusion of the loop-weight term $S_w$. The presence of this 
term in the effective action {\em fixes} the value of $K$, 
which in the traditional
Coulomb gas approach would remain unknown, only to be determined 
by an  exact value of some critical exponent, 
typically derived from a Bethe ansatz solution 
of the model. For the cases where Bethe ansatz solutions  
do exist (e.g., Potts models, $O(n)$ models)  
we have checked that the geometrical approach 
developed here reproduces the {\em same} value of the
stiffness (``renormalized coupling'') \cite{jkunp,jkfpl}.   

Microscopically, the operator $w({\bf h}({\bf x}))$ in $S_w$ 
generates the vertex weights  $\lambda({\bf x})$ in 
Fig.~\ref{vconf}d: $\lambda=\exp(-w)$. This operator can 
be written as 
$w({\bf h}({\bf x})) 
      = \frac{{\rm i}}{16} \:  {\bf E}_0 \cdot {\bf Q}({\bf x})$ 
\cite{jkunp},
where ${\bf Q}({\bf x})$ is the {\em cross-staggered} 
operator defined in Ref.~\cite{jkprb}; it is a vector valued 
function of the colors around the site ${\bf x}\in {\cal L}$. 
${\bf Q}({\bf x})$
is periodic in height space and it can be expanded in a Fourier
series of vertex operators 
$\exp({\rm i}{\bf G}\cdot{\bf h}({\bf x}))$; the vectors ${\bf G}$
lie in the lattice which is reciprocal to the lattice of height 
periods, and they are the electric charges in the Coulomb gas 
representation. 
The magnetic charges on the other hand form  the lattice
of height periods, and they are associated with 
vortex configurations
of the height with topological charge ${\bf b}$ \cite{jkprb}.   

In the long-wavelength limit 
we only keep the most {\em relevant} vertex operator appearing 
in the Fourier expansion of ${\bf Q}({\bf x})$; this  is 
the operator
with the smallest scaling dimension. The scaling dimension 
of a general operator with an  electro-magnetic charge 
$[{\bf G}, {\bf b}]$  is given by \cite{dots}:
\begin{equation}
\label{EMdims} 
x[{\bf G},{\bf b}] = \frac{1}{4 \pi K} \: {\bf G}\cdot({\bf G}-
 2{\bf E}_0) + \frac{K}{4 \pi} \: {\bf b}^2
\end{equation} 
Therefore, the loop weight term in Eq.~(\ref{action}) can 
be written as:   
\begin{equation}
\label{Sw}
  S_w = C e_0 \int \exp({\rm i}{\bf G}_Q \cdot {\bf h}({\bf x})) \: 
  d^2 {\bf x}   
\end{equation}
where $C$ is a constant independent of $n$, 
and ${\bf G}_Q=(-\pi,0,\pi)$ minimizes $x[{\bf G},0]$.

For $n\le 2$
the effective action $S$,   with $S_w$ given by
Eq.~(\ref{Sw}), 
describes a {\em Liouville field 
theory} with imaginary couplings;  $w({\bf h})$  is the so-called   
{\em Liouville potential}. In order for $S$ to describe a 
conformal field theory it is necessary that $w({\bf h})$
is a {\em marginal}  operator, i.e., $x[{\bf G}_{\bf Q},0]=2$ 
\cite{dots}. 
Moreover, if $S$ is a CFT  then 
$w({\bf h})$ is {\em exactly} marginal, and  the coupling 
$C e_0$ does not flow under renormalization. The exactly 
marginal operator $w({\bf h})$ gives rise
to a line of fixed points, which 
is the aforementioned critical phase of the loop model.  
For $n>2$ 
the parameter  $e_0$ in Eq.~(\ref{e0vsn}) becomes pure imaginary. 
From Eq.~(\ref{EMdims}), and the assumption that $K$ increases 
with $n$, it follows that in this case 
the scaling dimension of $w({\bf h})$
is necessarily less then two -- it is {\em relevant}. Therefore, 
under renormalization  
$n\to \infty$, and the loop model is in the ordered phase.

In the 
Coulomb gas representation of two-dimensional critical models 
the Liouville potential defines the 
so-called {\em screening charges}, 
which were originally  introduced on formal 
grounds -- to ensure the non-vanishing of the 
four-point correlation functions \cite{dots}. 
In the case of
loop models we have uncovered a geometrical interpretation of the 
Liouville potential:  
it enforces the correct weighting of oriented loops  by making the 
two orientations  inequivalent.

The assumption that $w({\bf h})$ is exactly marginal --
which we refer to as the {\em conformal ansatz} 
-- has a simple geometrical 
interpretation.  Namely, the loop weight ($n=2\cos(\pi e_0)$) is 
thermodynamically conjugate to the number of loops ($N$), 
and its non-renormalizability 
implies that the number of large loops, which are  
responsible for the long-wavelength fluctuations of the height,  
does not change under renormalization.
In this form the conformal ansatz  is 
closely analogous to the hypothesis often made for critical 
percolation, that  the number of spanning  
clusters is of order one, which in turn 
leads to hyperscaling \cite{coniglio}. 

\paragraph{Exact results}
The conformal ansatz implies $x[{\bf G}_{\bf Q},0]=2$, and  
from Eq.~(\ref{EMdims}) we calculate the exact value of the 
stiffness
\begin{equation}
\label{Kexact}
K = \frac{\pi}{4} \: (1 - e_0) \; ;
\end{equation}
for further convenience we define the parameter $g \equiv 4 K/\pi$. 
Using Eqs.~(\ref{e0vsn}) and~(\ref{Kexact}) for the coupling constants
of the effective field theory (Eq.~(\ref{action})), 
we can calculate the conformal charge and the critical exponents 
of the loop model along the critical line $n\le 2$. 

The conformal charge  is \cite{dots}:
\begin{equation}
\label{c-charge}
     c = 3 - 12 \: \frac{{\bf E}_0^2}{4 \pi K} = 3 - 12 \: 
                  \frac{(1-g)^2}{g} \; .
\end{equation} 
For $n=2$ and $n=1$, $g=1$ and $g=2/3$, and
consequently $c=3$ and $c=1$; $c=3$ is the conformal charge of the 
four-coloring model which can be mapped to the $n=2$ FPL${}^2$ model 
\cite{jkprb},
while $c=1$ is the known conformal charge of the equal-weighted 
six-vertex model \cite{cardyrev}, 
which is equivalent to the $n=1$ FPL${}^2$ model
\cite{squareFPL}. 

In the limit of Hamiltonian walks ($n\to 0$, $g=1/2$) 
we find $c=-3$. This differs from the 
conformal charge for  a single 
Hamiltonian  walk on the honeycomb ($c=-1$ \cite{batchfpl}) 
or the  square 
($c=-1.00(1)$ \cite{squareFPL})
 lattice, and we conclude that the FPL${}^2$
model defines a {\em new} 
universality class of self-avoiding random walks. 
This is also confirmed by the critical exponents, 
which we calculate next. 

The most commonly studied operators 
in the context of loop models are the 
so called string (``watermelon'') operators whose two-point 
correlation function gives 
the probability of having $m$ loop segments propagating between two 
points on the lattice \cite{dupl}; 
here we focus on string operators associated 
with black loops only.
In the Coulomb gas representation of the FPL${}^2$ model these 
operators are associated with mixed electric and magnetic charges:
$[(-e_0\pi/2, 0, e_0\pi/2), (-2k, 0, -2k)]$ is the charge for    
the $m=2k$ string operator, while 
$[(-e_0\pi, 0, 0), (-2k, -2, 2k-2)]$ for $m=2k-1$ \cite{jkunp}. 
These  charges can be
determined microscopically by mapping the string configurations
to vortex configurations of the height; 
an analogous analysis for the FPL model on the honeycomb lattice 
can be found in Ref.~\cite{jkfpl}. From Eq.~(\ref{EMdims}) we can 
calculate the  scaling dimensions of the string operators,    
\begin{eqnarray}
\label{exponents}
x_{2k} & = & \frac{g}{2} k^2 - \frac{(1-g)^2}{2g} \nonumber \\ 
x_{2k-1} & = & \frac{g}{2} (k^2 - k + 1) - \frac{(1-g)^2}{g} \; . 
\end{eqnarray}

The dimensions associated with an even number of strings are
the same as those found for the FPL model on the honeycomb lattice,
while the odd ones are new. In particular, in the Hamiltonian limit
($n\to 0$, $g=1/2$) we find $x_1=-1/4$. This, via standard scaling relations 
\cite{batchfpl}, leads to a {\em prediction} for the 
exponent $\gamma=1-x_1=5/4$, which describes
the scaling of the number of walks with the 
number of steps (bonds traversed). This value of 
$\gamma$ is different from the ones found
for a single walk on the honeycomb 
($\gamma=1$ \cite{batchfpl}) and the square lattice 
($\gamma\approx 1.0444$ \cite{squareFPL}), confirming once again 
that the FPL${}^2$ model defines a  new universality class 
of Hamiltonian  walks.

For $n=1$ the FPL${}^2$ model coincides with the FPL model of
Batchelor {\em et al.} \cite{squareFPL}.
For this value of $n$, $g=2/3$, and  
$x_1=1/6$ and  $x_2=1/4$, while the numerical 
transfer matrix results quoted in Ref.\cite{squareFPL} 
are $x_1=0.1667(1)$ and $x_2=0.2500(1)$.
Furthermore Batchelor {\em et al.}
found numerical evidence for
the relation $x_3 = x_1 + g$, which we find to be satisfied exactly.

In conclusion, we have shown that the conformal field theory 
associated with the scaling limit of a lattice loop-model 
can be constructed by mapping it to a random surface. The loops 
appear as contour lines of a  surface whose long wavelength 
fluctuations are described by a Liouville field theory. 
The remarkable feature of loop models is that the couplings
in the field theory are completely fixed by geometry thus 
leading to exact results for their critical properties.

The conformal ansatz, which is equivalent to the statement that 
the number of large loops does not change with  length scale, 
is central to the whole construction, and it leads to a physical 
interpretation of the Dotsenko-Fateev screening charges \cite{dots}. 
The validity of this ansatz has been confirmed for loop models 
for which an exact 
solution exists (e.g., Potts models, $O(n)$ models) 
\cite{jkunp,jkfpl}. In order to show  this   
explicitly a  renormalization  
group which takes into account the extended nature of loops 
will have to be developed. This remains an interesting open question. 

Enlightening  discussions with J. Cardy, J. deGier, 
C.L. Henley, G. Huber, 
J.B. Marston, B. Nienhuis, V. Pasquier  and T. Spencer 
are acknowledged. 
I am particularly 
indebted to J. Cardy who alerted me to  the relation between 
the Liouville potential and the screening charges, and to C.L. Henley 
for pointing out 
the relation between the loop weight and the number of large loops. 
It is also a pleasure to acknowledge the
hospitality of T. Spencer and 
the Institute for Advanced Study, where part of this work was 
completed. This work was supported by the NSF
through grant No.~DMR-9357613.

\end{multicols}

\end{document}